\documentclass[twocolumn,english,aps,PRA,reprint, superscriptaddress,showpacs,longbibliography,showkeys]{revtex4-1}
\usepackage{amsmath,amssymb,bbm,mathrsfs,bm,braket,color,graphicx,comment}
\usepackage[colorlinks,citecolor=blue,urlcolor=blue,linkcolor = blue]{hyperref}
\usepackage[mathscr]{euscript}

\usepackage{xcolor}
\newcommand\scalemath[2]{\scalebox{#1}{\mbox{\ensuremath{\displaystyle #2}}}}

\begin{document}

\title{Minimal Constraints in the Parity Formulation of Optimization Problems}
\author{Martin Lanthaler}
\email{Martin.Lanthaler@uibk.ac.at} 
\affiliation{Institute for Theoretical Physics, University of Innsbruck, A-6020 Innsbruck, Austria}
\author{Wolfgang Lechner}
\email{Wolfgang.Lechner@uibk.ac.at}
\affiliation{Institute for Theoretical Physics, University of Innsbruck, A-6020 Innsbruck, Austria}
\affiliation{Parity Quantum Computing GmbH, A-6020 Innsbruck, Austria}

\begin{abstract}
As a means to solve optimization problems using quantum computers, the problem is typically recast into a Ising spin model whose ground-state is the solution of the optimization problem. An alternative to the Ising formulation is the Lechner-Hauke-Zoller model, which has the form of a lattice gauge model with nearest neighbor 4-body constraints. Here we introduce a method to find the minimal strength of the constraints which are required to conserve the correct ground-state. Based on this, we derive upper and lower bounds for the minimal constraints strengths. We find that depending on the problem class, the exponent ranges from linear $\alpha \propto 1$ to quadratic $\alpha \propto 2$ scaling with the number of logical qubits.

\end{abstract}
\pacs{}
\maketitle

\section{Introduction}
\label{sec:Intro}

Combinatorial optimization problems are ubiquitous in a wide range of scientific fields. 
Most of these problems can be reformulated as an Ising-spin glass problem \cite{Lucas}, which is the starting point for digital (e.g.\@ quantum approximate optimization algorithm QAOA) \cite{FarhiQAOA} and analog quantum optimization algorithms (e.g.\@ adiabatic quantum optimization AQO) \cite{FGGS,SusaNishimori,Kadowaki1998QA,Kadowaki2002Thesis}.
The goal of both approaches is to find low energy states of the spin model, which correspond to solutions of the previously encoded optimization problem. The efficiency of AQO, in particular the claim of substantial speedup is currently under debate (for review see \cite{AlbLid} or \cite{Hauke2020PerspectivesOQ}). It is not ruled out that highly coherent AQO may be more efficient than classical algorithms, at least for some classes of problems \cite{Katzgraber2015SeekingQS,LidarTroyerGoogleQSU,MbengSQA19,SantoroAnnealing,Dick,HartmannCD,HartmannIHD}. In QAOA, low energy states are found via a variational procedure which is considered a promising route for near term quantum optimization \cite{ZhouPichlerLukin2020QAOA,Farhi2016QSuprem,Wecker2016}. However, in both, adiabatic and digital algorithms, the problem Hamiltonian contains long-range interactions which requires either embedding schemes \cite{Choi2008ME,Choi2010ME, Bunyk2014Chimera, LHZ} or large numbers of SWAP operations. 

\begin{figure}
        \includegraphics[width = .5\textwidth]{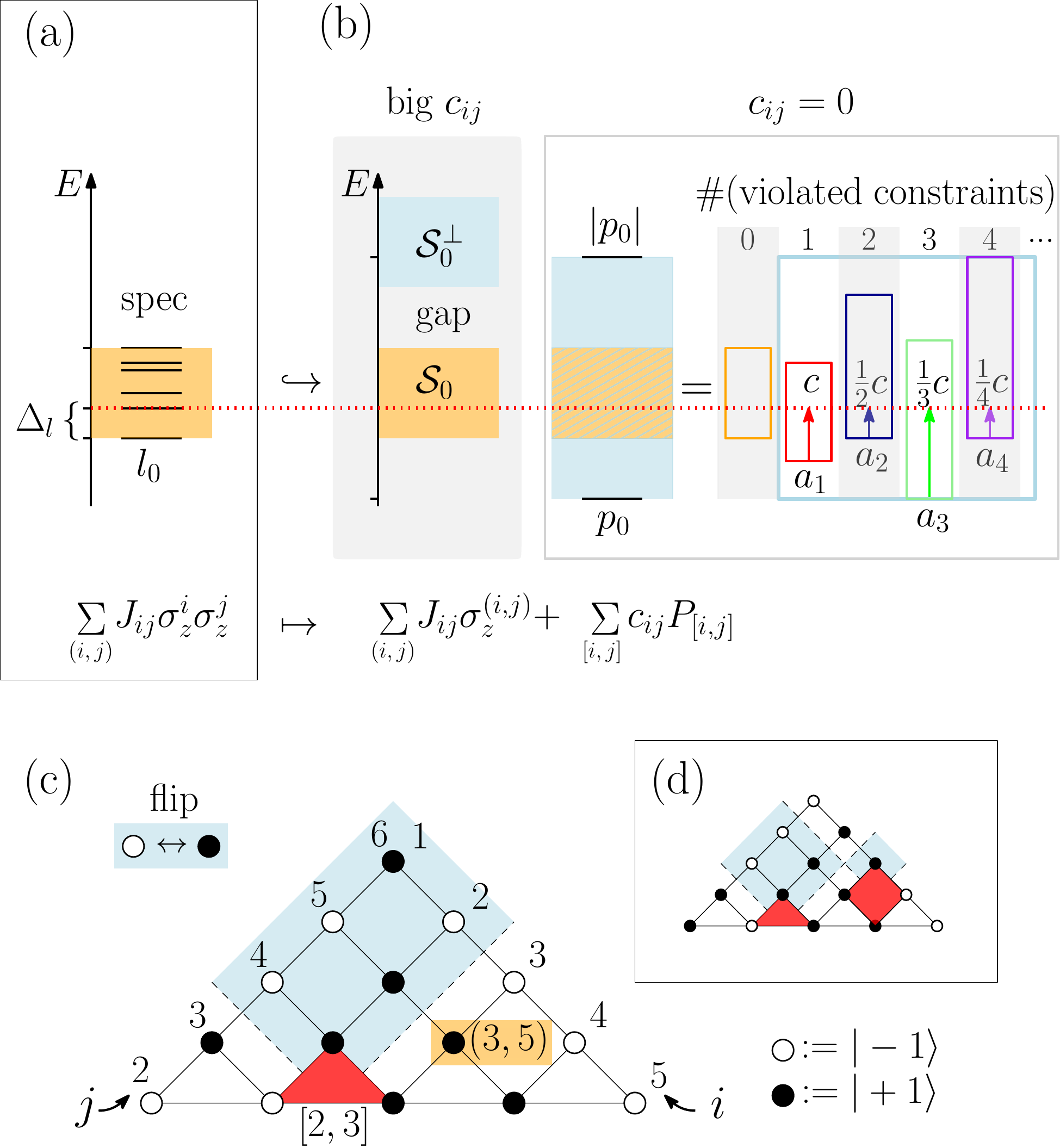}
        \caption[l]{
            (a,b) Following the parity architecture \cite{LHZ}, the logical spectra can be found (up to a global shift) as a subset the parity Hamiltonians spectra. 
            To ensure a separation, local 4-body terms are introduced via penalty terms $P_{[i,j]} = \sigma_z^{(i,j)}\sigma_z^{(i,j+1)}\sigma_z^{(i+1,j)}\sigma_z^{(i+1,j+1)}/2$, as denoted in  Eq.~\eqref{eq:ParityHamiltonian}.
            If the corresponding strengths $c_{ij}$ are chosen large enough, the original spectra is well separated from all the other eigenenergies (b , left). 
            The required strengths can be lowered by allowing  unwanted energy levels as low as the the first excited state $e = l_0+\Delta_l$.
            For $c_{ij} = c$, finding the minimal strength $c$ involves minimizing over subspaces with defined number of violated constraints and then taking the largest $c$. A systematic scheme to construct violating states is shown in panel (c) and (d). Starting from a state with no parity constraint violated and flipping the spins in the blue shaded region, one can construct all states with certain parity constraints violated.
        }
        \label{fig:Parity}
\end{figure}

An alternative to the spin glass paradigm has been recently introduced \cite{LHZ,LechnerQAOA,GvBZL,Leib2016Transmon}. 
By a conceptual division of logical qubits, defining the optimization problem and the physical qubits available in the laboratory one maps the logical Ising Hamiltonian
$H_{\mathrm{logic}} = \sum_{(i,j)} J_{ij} \sigma_z^i  \sigma_z^j$
to the physical Hamiltonian
\begin{equation}
\begin{split}
    H_\mathrm{phys} &= \sum_{(i,j)} J_{ij} \sigma_z^{(i,j)}  \\
        &- \frac{1}{2}\sum_{[i,j]}
    c_{ij}\sigma_z^{(i,j)}\sigma_z^{(i,j+1)}\sigma_z^{(i+1,j)}\sigma_z^{(i+1,j+1)} .
\end{split}
\label{eq:ParityHamiltonian}
\end{equation}
This mapping is done by introducing a physical qubit for each pair of logical qubits, where the z-component corresponds to the relative orientation of two logical qubits  i.e.\@ $\sigma_z^{(i,j)} := \sigma_z^i \sigma_z^j$.
The overhead in qubits for all-to-all pair interactions is quadratic and this increased number of degrees of freedom is compensated by constraints. Arranging the physical spins on a 2D lattice, allows to construct the constraints from 4-local interaction on individual plaquettes consisting of 4 neighbouring spins. Figure~\ref{fig:Parity} sketches the layout and the labeling of the physical qubits $(i,j)$ and plaquettes $[i,j]$, where the labels only run over pairs with $i < j$. 
The 4-body constraints ensure that the low energy sub-space of the physical system $H_{\mathrm{phys}}$ is exactly the spectra of the logical Hamiltonian $H_{\mathrm{logic}}$ by adjusting the
constraint strengths $c_{ij} \in \mathbb{R}$.  The constraints $c_{ij}$ have to be chosen large enough to separate the allowed \emph{logical subspace}, i.e.\@ 
states which do have a translation back into the logical picture, from states which do not have a counterpart in the logical model. The scaling of the constraint strengths is also crucial for the performance of the quantum annealing protocol \cite{AlbLidSQA}.

In this paper, we determine the minimum constraint strengths $\hat{c}_{ij}$ that satisfy the lowest and first exited state of the problem Hamiltonian. In favour to reduce the magnitude of the constraint strengths we drop the requirement for a full separation between logical sub-spectra and the other eigenvalues. 
We show that finding the minimal constraints can be rewritten as a linear program. In the homogeneous setting $c_{ij} = c$, we derive a series of upper and lower bounds to the optimal constraint strength $c$ allowing to approximate the optimal values. 
Different classes of optimization problems are modelled by considering $J_{ij}$ as independent and identically distributed  (i.i.d.\@) random variables 
with probability density function (pdf) $f(\mu,\sigma^2)$.
In the case $\mu/\sigma \to \pm \infty$ we derive analytic solutions to the minimal constraint problem, which are naturally related to the problem of solving \textsc{MaxCut} on the complete graph $K_n$ or the total ferromagnetic problem respectively. By a simple argument the authors in \cite{AlbLidSQA} concluded, that in the antiferromagnetic case the constraints should grow at least linearly with the system size. We show, that in this case the constraints even have to grow quadratic in system size.
Also for random $J_{ij}\in \{-1,1\}$ the authors of \cite{AlbLidSQA} expect the constraints to scale linearly with the size of the problem.
We find, for the case of $\mu/\sigma$ finite, the large size scaling of the expected optimal constraint strength is mainly determined by the sign of the expectation value $\mu$. If $\mu$ is negative, the large size scaling becomes linear. Furthermore, in the case $\mu$ positive the scaling becomes quadratic. The point $\mu = 0$ is interesting for symmetry reasons.  By relying on results from extreme value theory we argue, that for standard Gaussian couplings,  choosing the constraint strengths of order $\sqrt{n}\log(n)$ could be enough to ensure that the physical ground state faithfully represents the logical ground state. 

\section{Constraints}

The constraints separate the subspace of allowed configurations from the unphysical subspace. The strength of the constraints have to be large compared to the energy of the local field energies in the system. In the following we derive both, upper and lower bounds for the constraint strength. 

\subsection{Minimal constraint problem}

Our goal is to find the minimal strength of the constraints, such that lowest and first excited states
of the physical and logical Hamiltonian coincide w.r.t.\@ the parity translation. 
Intuitively, this can be understood as follows: In the extreme case, where the constraints are set to zero, each spin would point in the direction of the local field acting on the spin in order to minimize the energy of the system. On the contrary, if the constraints are infinitely large, these states are generally forbidden and the condition of an even number of spins up per plaquette is enforced. 

We consider the case of finite constraints where the local field term and the  constraint energies are competing. In this case it might be energetically favorable to violate a constraint for rearranging the spins with respect to their local fields. Our goal is to to find the lowest constraint energy such that this case can be ruled out. The minimal energy and the scaling w.r.t.\@ the number of qubits depends on the statistics of the local fields which in turn is associated with classes of optimization problems. Therefore, we derive the minimal constraints for different classes of optimization problems.  

We consider $n$ logical spins $[n]:=\{1,..,n\}$ with all-to-all connectivity. Thus, the system contains $m := n(n-1)/2$ interactions that are mapped to $m$ spins in the parity scheme.
The interaction strengths can be viewed as weights on the edges of a complete Graph $K_n = (V_n,E_n)$ with $V_n :=[n]$ and $E_n := \{(i,j)\in [n]^2, i<j\}$.  We label the physical spins with elements of $E_n$ 
[cf.\@ Fig.~\ref{fig:Parity}(c)]. Plaquettes are labeled by elements from $E_{n-1}$ and to distinguish them from sites, we replace the curly brackets $(\;)$ with square ones $[\;]$.
Furthermore, we denote the sample mean of a random variable $X$ by $\overline{X}$.

The space of physical states $\{-1,1\}^m$ can be decomposed into a family of subspaces $(\mathcal{S}_{\omega})_{\omega \subseteq E_{n-1}}$ - according to their pattern on individual plaquettes.
In that sense $\mathcal{S}_0 := \mathcal{S}_{\{\}}$ should denote the logical subspace where the local constraints on every plaquette are satisfied.
More general, given a tuple $\omega \subseteq  E_{n-1}$ of plaquettes, the subspaces $\mathcal{S}_k$ are defined as
states being simultaneous eigenstates to the stabilizers $P_{[i,j]} = -\sigma_z^{(i,j)}\sigma_z^{(i,j+1)}\sigma_z^{(i+1,j)}\sigma_z^{(i+1,j+1)}/2$ with corresponding eigenvalues $\frac{1}{2}$ if $(i,j) \in \omega$ and $-\frac{1}{2}$ otherwise. 
Note, on the lowest row of plaquettes indexed with $j = i+1$, the stabilizers $P_{[i,j]}$ are given by 3-local terms $ \sigma_z^{(i,i+1)}\sigma_z^{(i,i+2)}\sigma_z^{(i+1,i+2)}/2$.
In Fig.~\ref{fig:Parity}(c)  we show an example for a state satisfying all constraints beside the one corresponding to the loop $23 - 34 - 42$. That state belongs to in the subspace $\mathcal{S}_{[2,3]}$. Likewise Fig.~\ref{fig:Parity}(d) shows another state with two unsatisfied constraints i.e.\@ being element of $\mathcal{S}_{\{[2,3],[3,5]\}}$.\\

The constraint strengths $c_{ij}$ can be chosen either to be all identical (homogeneous case) or we can individually set them to the optimal constraint strength for each plaquette. In the latter case, the objective cost function depends on all individual constraints $\mathrm{cost}(c_{12},...,c_{n-1,n})$, which in the linear case is the sum of the constraint strengths. 
Let us write $H_{\mathrm{phys}} = H_J+H_c$, with $H_J$ the part of the physical Hamiltonian related to the local fields and $H_c$ the term related to the constraints.
We define the value $a_{\omega}$ as the lowest eigenenergy of $H_J$ restricted to states belonging to the subspace $\mathcal{S}_{\omega}$.
With this definition, the 
problem of minimizing the constraints strengths can be written as a linear program: To this end, the cost-function has to be minimized under the restrictions
\begin{equation}
\label{eq:LinearProgram}
    \sum_{(i,j)\in \omega} c_{ij}  \geq -a_{\omega} + e, \qquad \forall \omega \subseteq E_{n-1}, \omega \neq \emptyset
    ,
\end{equation}
where the value $e$ denotes the first exited eigenenergy of the problem Hamiltonian.

In the homogeneous case $c_{ij} = c$, 
the linear program Eq. ~\eqref{eq:LinearProgram} reduces to
\begin{equation}
\label{eq:HomogeneousConstraint}
    c = \max\left\{e-a_1,\frac{1}{2}(e-a_2),...,\frac{1}{q}(e-a_q) \right\},
\end{equation}
were $a_k := \min\{a_{\omega}: \omega \subseteq E_{n-1}, |\omega|=k\}$
and $q$ denotes the number of plaquettes.
Here, the penalty Hamiltonian $H_c$ does not discriminate between states with the same number of parity constraints violated. 
Since $a_1$ is the lowest eigenenergy of $H_J$ w.r.t.\@ the subspace
$\mathcal{S}_1 := \mathcal{S}_{[1,2]} \cup ... \cup \mathcal{S}_{[n-2,n-1]}$,
where a single parity condition is unsatisfied,
the corresponding state gets a penalty of $c$.  This penalty has to be chosen large enough to bridge the gap between $a_1$ and $e$, which explains the first term in Eq.~\eqref{eq:HomogeneousConstraint}.  Similar to $\mathcal{S}_1$ we define $\mathcal{S}_2$ as the subspace of states with two unsatisfied parity constraints. $a_2$ is then given as the lowest eigenvalue of $H_J$ restricted to $\mathcal{S}_2$. 
Since all these states will penalised twice by $H_c$, the strength of $c$ has to be at least half the difference of $a_2$ and $e$. This explains the second term in Eq.~\eqref{eq:HomogeneousConstraint}. Finally, other cases with $k>2$ follow by including states with more than two unsatisfied parity constraints.

In general, every term appearing in Eq.~\eqref{eq:HomogeneousConstraint} is of the form $c_{-k}:= (e-a_k)/k$
and can be seen as a lower bound  for the optimal constraint strength.
To get upper bounds, we consider the fact that the spectrum of $H_J$ is contained in the interval $[p_0,-p_0]$ with boundaries $p_0 := -\sum_{i<j} |J_{ij}|$. With the definition 
\begin{equation}
\label{eq:HomogeneousUpperBounds}
    c_i := \max \left\{ c_{-1},c_{-2},...,c_{-i},\frac{1}{i+1}(e-p_0)\right\}
\end{equation}
a series of upper bounds can be derived according to
\begin{equation}
\label{eq:HomogeneousBoundsChain}
c \leq c_q \leq c_{q-1} \leq ... \leq c_1 \leq c_0 \leq 2|p_0| . 
\end{equation}
Note, that we included the trivial bound $2|p_0|$ and defined $c_0 := e-p_0$ in Eq.~\eqref{eq:HomogeneousBoundsChain}.\\

\subsection{Single violator approximation}

In order to make the problem numerically more accessible, we focus on the first lower bound $c_{-1}$ rather than $c$. Thus, only states with one parity constraints violated are considered. We like to call them the \emph{single violators} states [cf.\@ Fig.~\ref{fig:Parity}(c)]. 
This is numerically well justified since for all models studied in this manuscript we observe the ordering
\begin{equation}
    \overline{c} \approx \overline{c_{-1}}\geq \overline{c_{-2}}\geq \cdots ,
\end{equation}
[cf.\@ Fig.~\ref{fig:SKSpinGlassNumerics} and Fig.~\ref{fig:MaxCutNumerics}].
\\

It is easy to see, if the $J_{ij}$ are $m$ i.i.d.\@ random variables, the expected minimal constraint strength $\overline{c}$ cannot grow faster than quadratic in $n$, since by the central limit theorem we have
\begin{equation}
   2 |p_0| \to N(2m \mu_{\mathrm{abs}} ,\sigma_{\mathrm{abs}}^2 2 m)
\end{equation}
for $n \to \infty$, where $N(\mu,\sigma)$ denotes the normal distribution and $\mu_{\mathrm{abs}}$ and $\sigma_{\mathrm{abs}}^2$ are the mean and variance of the positive random variables $|J_{ij}|$.
Therefore, the trivial upper bound scales quadratic $2\overline{|p_0|}  = \Theta(n^2)$, and with $\overline{c} \leq 2\overline{|p_0|}$ one further concludes that the minimal constraint strength cannot grow faster than quadratic in $n$ i.e.\@ $\overline{c} = \mathcal{O}(n^2)$.

Furthermore, if $J_{ij}$ are i.i.d.\@ random variables, with pdf $f_{\mu,\sigma^2}(x)$, the problem of determining the scaling of $\overline{c}$ does only depends on the ratio $\mu/\sigma$.
This can be seen by noting that a rescaling of the pdf $f(x) \mapsto f(k^{-1} x)$ is equivalent to multiplying the random variables by a constant factor $J_{ij} \mapsto k J_{ij}$. 
Hence, the strengths of the optimal constraints are multiplied by an overall factor of $k$ whereas the functional dependency on the size, i.e.\@ the scaling of the optimal constraints, is not affected.
On the other hand, for each random variable it is true that $\overline{(k J_{ij})} = k \overline{J_{ij}}$ and $\mathrm{var}(k J_{ij}) = k^2 \mathrm{var}(J_{ij})$ i.e.\@ rescaling of $J_{ij}$ does not alter $\mu/\sigma$. 
In conclusion, the scaling of the optimal constraints can only depend on the ratio  $\mu/\sigma$.

\section{Results}

Using the bounds Eq.~\eqref{eq:HomogeneousUpperBounds} we evaluate the optimal constraints, for general ensembles of systems with different specific connectivity, bias and variance. In particular, the scaling of the average optimal constraint strength $\overline{c}$ with the system size for classes of problems. Let us first introduce two examples for typical optimization problems.\\

Let $G=(E,V)$ denote a simple graph.
Then, the \textsc{MaxCut} problem asks for two disjoint sets of vertices $V_1$ and $V_2$ with  $V_1 \cup V_2 = V$, such that the number of cutting edges is maximal i.e.\@ $(e_1,e_2)$ with $e_1 \in V_1$ and $e_2 \in V_2$. As second example, the \textsc{MinBisection} (or graph-bipartitioning) problem for a graph with even number of nodes, requires to minimize the number of cutting edges while balancing the size of the two subsets $|V_1|=|V_2|= |V|/2$.

These graph partitioning problems can be easily mapped onto an Ising problem by introducing one spin per node. 
The \textsc{MaxCut} problem can be reformulated as an \emph{antiferromagnetic} Ising model, i.e.\@ $J_{ij} = 1$ for
all $(i,j)\in E$, where the ground state corresponds to the solution of the optimization problem. 
If $l_0$ denotes the smallest eigenvalue of 
\begin{equation}
    H_\mathrm{\textsc{MaxCut}} = \sum_{(i,j)\in E}\sigma_z^i\sigma_z^j,
    \label{eq:MaxCutIsingFormulation}
\end{equation}
then the maximal cut is given by 
$\mathrm{cut}_{\max} =(-l_0 + |E|)/2$.

Similarly, the \textsc{MinBisection} problem  can be encoded into an \emph{ferromagnetic} Ising model with magnetization fixed to zero, i.e.\@ $J_{ij} = -1$ for all $(i,j)\in E$, with $\sum \sigma_z^i = 0$. The corresponding Hamiltonian reads as
\begin{equation}
    H_\mathrm{\textsc{MinBisection}} = -\sum_{(i,j)\in E}\sigma_z^i\sigma_z^j + u\left(\sum_i \sigma_z^i\right)^2,
    \label{eq:MinBisectionIsing}
\end{equation}
where the second term of Eq.~\eqref{eq:MinBisectionIsing} guarantees, that the magnetization of the ground state is zero, given the energy penalty $u$ is larger than $\min(4 d_{\max},n)/4$, with $d_{\max}$ the maximal degree of $G$ \cite{Lucas}.

\begin{figure}
        \includegraphics[width = .5\textwidth]{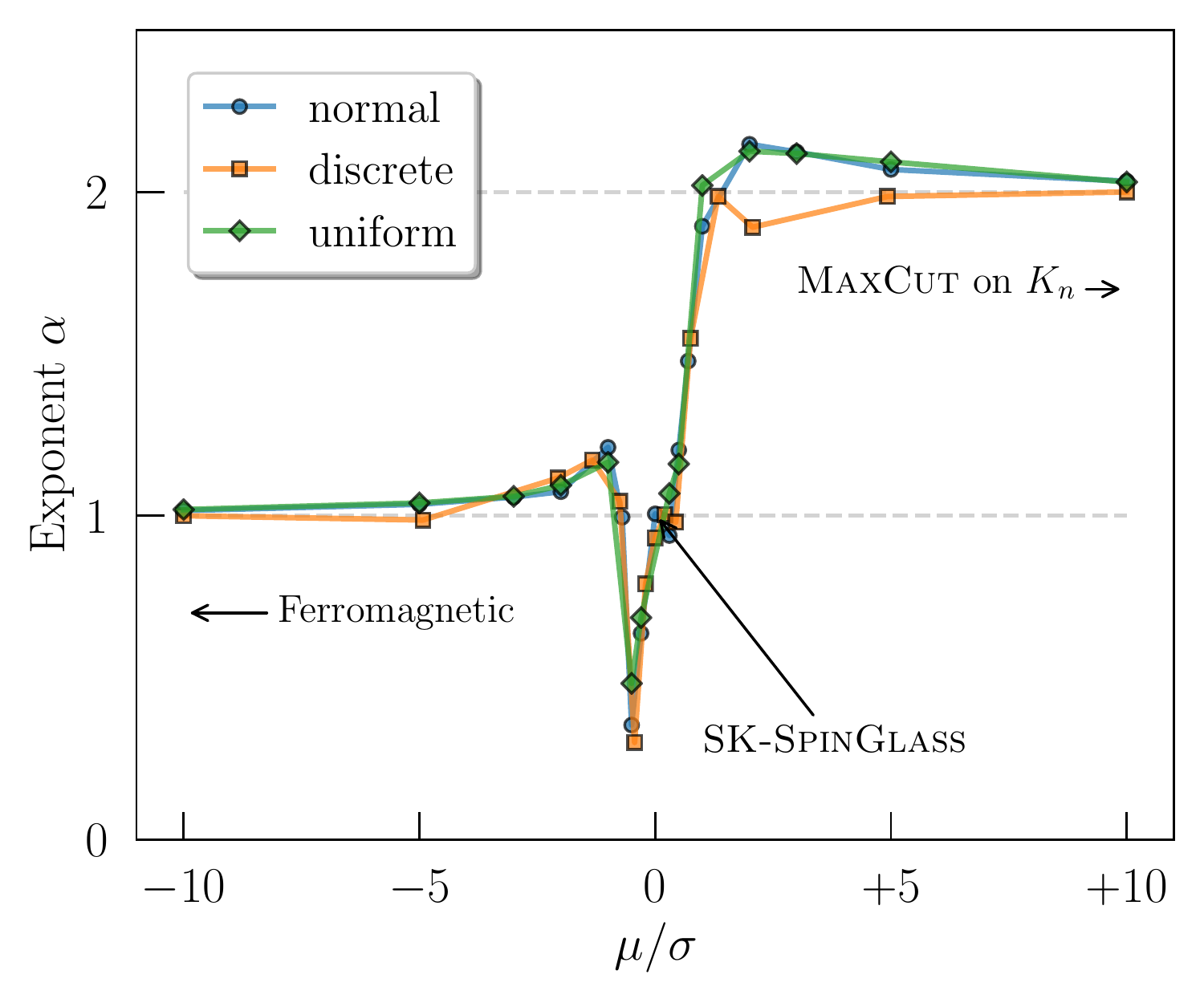}
        \caption[l]{
        Scaling of the minimal constraint strengths $\overline{c}$ 
        for the particular cases when the coupling strengths $J_{ij}$ are chosen normally, uniformly or according to a bi-modal discrete distribution with expectation value $\mu$ and variance $\sigma^2$.
        The plot shows the exponent $\alpha$ obtained from the power-law fit $n \mapsto \beta n^{\alpha} +\gamma$ onto the lower bound $\overline{c_{-1}}$ for simulations up to system sizes of $n = 25$.
        }
        \label{fig:Slopes_mu_over_sigma}
\end{figure}

\subsection{Numerical Results}
\label{sec:NumericalResults}

The general case of randomly distributed $J_{ij}$ values with a given bias $\mu$ and standard deviation $\sigma$ can be treated numerically. In the following we investigate and compare three different distributions.

1.) Normal distribution $N(\mu,\sigma)$ with mean $\mu$ and variance $\sigma^2$. 

2.) Uniform distribution on the interval $[a,b]$ with $\mu/\sigma = \sqrt{3}(a+b)/|a-b|$. 

3.) A bimodal distribution with two possible assignments $\{-1,1\}$ where the probability to choose $+1$ equals $p$ and $\mu/\sigma = (2p-1)/2[p(1-p)]^{\frac{1}{2}}$.\\

\begin{figure}
        \includegraphics[width = .5\textwidth]{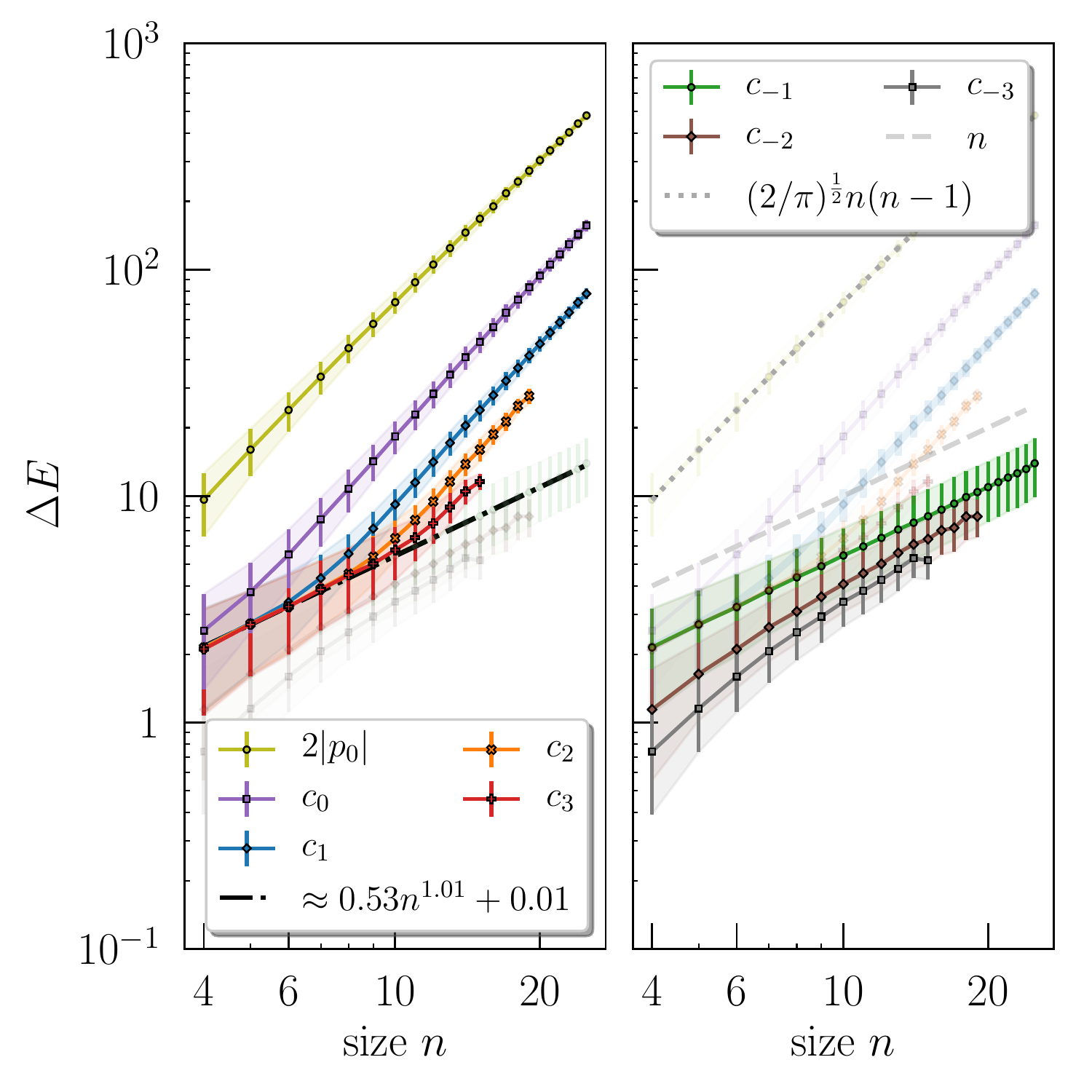}
        \caption[p]{
        Numerical simulation for random instances for \textsc{SK-SpinGlass}  problems i.e.\@ $J_{ij}$ i.i.d.\@ $\sim N(0,1)$.
        The log-log-plot spans the logical system size from $n=4$ to $n=25$. The corresponding number of instances drop from $10^5$ for $n=4$ to $64$ for $n=25$.
        Shown are mean and variance for upper bounds (left) and for the lower bounds (right). The dash-dotted curve is the result from the fitting model $n \mapsto \beta n^{\alpha} + \gamma$ applied on $\overline{c_{-1}}$.
        }
        \label{fig:SKSpinGlassNumerics}
\end{figure}

\begin{figure*}
        \includegraphics[width =\textwidth]{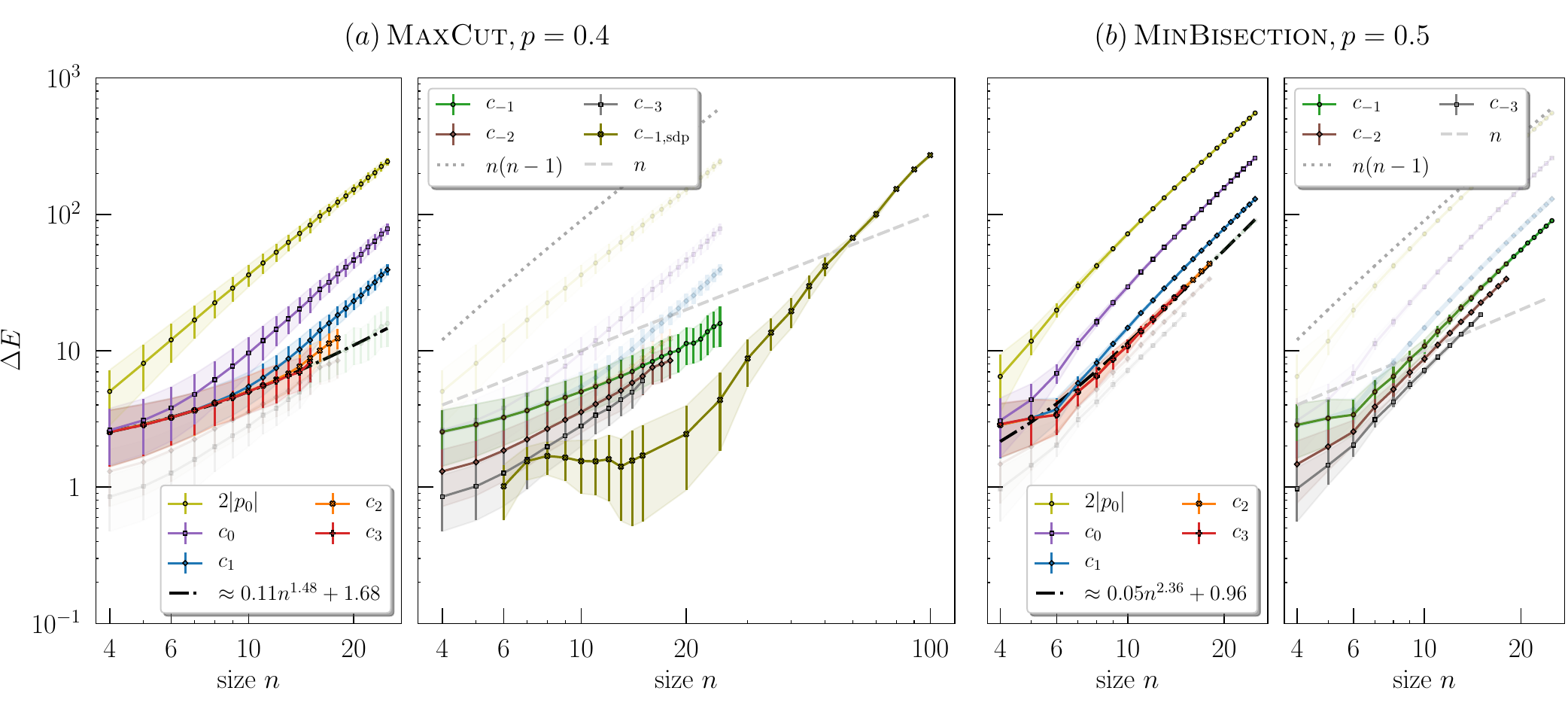}
        \caption[n]{(a) Numerical simulations of \textsc{MaxCut} instances based on random Erd\H{o}s-Réni graphs with $p=0.4$. Shown are variance and mean of upper bounds (left) and lower bounds (right) to the minimal constraint strengths. Furthermore, building on semidefinite programming, an efficient calculable lower bound is provided. 
        (b) Numerical results for \textsc{MaxBisection} instances on random graphs with $p=0.5$ for upper and lower bound respectively.
        }
        \label{fig:MaxCutNumerics}
\end{figure*}

For our numerical results we sample from random instances of particular optimization problems, calculate their ground state and lowest single violator energies and interpolate the sample mean with a powerlaw fit. Finding $a_1$ involves minimisation over subspaces with a single parity defect.
These single violator states we enumerate them by flipping spins starting from a state from the logical subspace [cf.\@ Fig.~\ref{fig:Parity}(c)]. This allows us to do reasonable statistics up to sizes of $n=25$. In the parity picture this corresponds to $m=300$ physical spins and $q = 276$ plaquettes.

Assuming i.i.d.\@ random variables $J_{ij} = J$,  homogeneous constraint strengths $c_{ij} = c$
and single violator approximation $\overline{c}\approx \overline{c_{-1}}$, the numerics in  Figure~\ref{fig:Slopes_mu_over_sigma} suggests that the scaling of the minimal constraint strength mainly depends on $\mu$, the expectation value of $J$. 
The limits $\mu/\sigma\to \pm \infty$ are analytically well understood, showing a linear and a quadratic scaling respectively. Furthermore, at $\mu/\sigma = 0$ all three analyzed distributions show a linear scaling, including the \textsc{SK-SpinGlass} model where $J_{ij}$ are standard normal distributed random variables [cf.\@ Fig.~\ref{fig:SKSpinGlassNumerics}]. Note, that this behaviour may only occur in small systems, as we will further elaborate in section~\ref{sec:AnalyticalResults}(c). \\

As paradigmatic examples of combinatorial optimization problems - satisfying the assumption of independent random variables - we consider now two graph partitioning problems on random Erdős–Rényi graphs, where each edge has a fixed probability $p$ of being present. \\

\paragraph{\textsc{MaxCut}:}
Solving \textsc{MaxCut} for random graphs, corresponds to independently choosing $J_{ij}$ with probability $p$ to be either $1$ or $0$, respectively.
Since all strengths fulfill $J_{ij}\geq0$, it follows that $\mu/\sigma \geq 0$.
Fig.~\ref{fig:MaxCutNumerics} shows our numerical results for system sizes up to $n=25$, where we find a sub-quadratic increase of $\overline{c_{-1}}$ with the system size. As we will see in the following analysis, this sub-quadratic scaling becomes quadratic for larger problem sizes. 
To this end we derive an efficiently calculable lower bound by utilizing a semidefinite program relaxation of finding the ground state configuration of Hamiltonian~\eqref{eq:MaxCutIsingFormulation} with $\sigma^i_z\in \{-1,1\}$.
If $\mathrm{opt}_{\mathrm{sdp}}$ denotes the optimal value of the semidefinite program
\begin{equation}
    \max \sum_{(i,j)\in E}\frac{1 - X_{ij}}{2},
    \qquad 
    X_{ii} = 1\; 
    \forall i\in [n], X \succeq 0
\end{equation} 
then $\mathrm{cut}_{\max} \leq \mathrm{opt}_{\mathrm{sdp}}$ \cite{GoemansWilliamson1995improved}.

To be computationally efficient, we upper bound the minimal single-violator contribution $a_1 \leq a_1^+$ by sampling from quadratically many single-defect states. To achieve this, the set of vertices $V = [n]$ 
is partitioned into three disjoint sets containing consecutive nodes
$A = \{1,2,...,k\}$,
$B = \{k+1,...j\}$ and 
$C = \{j+1,...,n\}$,
where all sets contain at least two elements. The lower right part of Fig.~\ref{fig:MAXMINCut}, demonstrates what such a partition of  $n=6$ nodes into sets $A = \{1,2\}$, $B = \{3,4\}$ and $C= \{5,6\}$ looks like in the parity picture. Note, that this particular state has a single unsatisfied constraint at plaquette $[2,4]$. With $l_0 +2 \leq l_2$ and $\mathrm{cut}_{\max} = (-l_0+|E|)/2$ we can derive a lower bound on $c_{-1} = l_2-a_1\geq c_{-1,\mathrm{sdp}} $ by defining
\begin{equation}
    c_{-1,\mathrm{sdp}} := -2\cdot \mathrm{opt}_{\mathrm{sdp}} + |E| + 2 - a_1^{+}.
\end{equation}
Figure~\ref{fig:MaxCutNumerics} includes this lower bound $c_{-1,\mathrm{sdp}}$ for the class of random graphs with edge probability $p=0.4$ up to system sizes of $n=100$. We find, that after a sub-quadratic increase, the growth rate of $\overline{c}$ becomes quadratic for large sizes $n$.

\paragraph{\textsc{MinBisection}:}
It turns out, that the special case of \textsc{MaxCut} problems on the complete graph $K_n$, is key to understand the behaviour of the minimal constraints in the \textsc{MinBisection} problem for large system sizes $n$. 
For fixed $p$,  $u$ scales at least linearly in  $n$. Hence, the inverse $u^{-1}$  faster approaches zero than $n^{-1}$. This means, if $n$ is large enough
\begin{equation}
\label{eq:MinBisectionIsing}
\frac{H_{\mathrm{\textsc{MinBisection}}}}{u} \propto 
\sum_{(i,j) \in E_n}\sigma_z^i\sigma_z^j - \frac{1}{u} \sum_{(i,j)\in E}\sigma_z^i\sigma_z^j,
\end{equation}
is well approximated by the first term $\sum_{(i,j)\in E_n}\sigma_z^i\sigma_z^j$, i.e.\@ equals the Hamiltonian for \textsc{MaxCut} problems, when restricted to the class of complete graphs $K_n$. 
Therefore, for  large $n$, the expected value of $c_{-1}$ grows quadratic, independent of the choice of $p$, as we will further elaborate in the subsequent section [cf.\@Fig~\ref{fig:MaxCutNumerics}(b)].\\

\subsection{Analytical Results}
\label{sec:AnalyticalResults}
Now we investigate two limiting cases ($\mu/\sigma \to \pm \infty$) where the minimal constraint problem can be solved analytically. Furthermore, we discuss the case of normally distributed interaction strengths with $\mu = 0$ and $\sigma \neq 0$.\\

\begin{figure}
        \includegraphics[width = .5\textwidth]{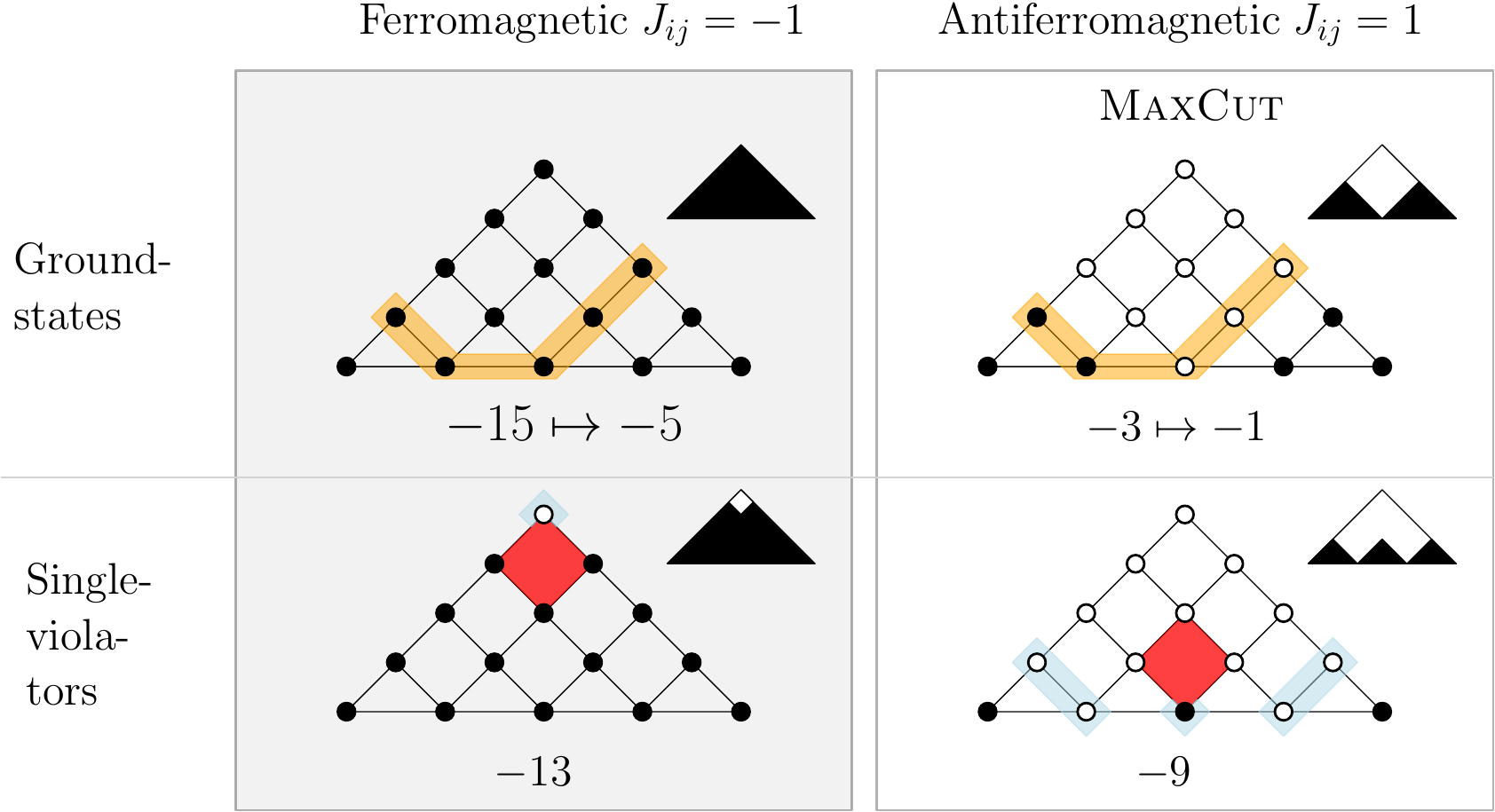}
        \caption[l]{ 
        Fully ferromagnetic and anti-ferromagnetic ground states and single violator states.
        The ground state of the ferromagnetic Hamiltonian, corresponds to collecting all vertices in the same set. 
        Missing a single one gives the first exited state, which hurts $n-1$ of the interactions.
        Flipping the spin $(1,n)$ introduces a defect, which comes with a energy cost of $2$.
        For $n$ even, a solution to \textsc{MaxCut} split the set of vertices into two subsets of equal size.
        The difference between cutting edges (white circles) and in-set edges (black circles) grow linear $l_0 = -n/2$. 
        The first exited state is obtained by splitting into sets with $\frac{n}{2}-1$ and $\frac{n}{2}+1$ elements respectively, which gives an additional linear energy shift of $2(n-1)$.
        The ground state single violator states are obtained by collecting the vertices into three sets, i.e.\@ for $n= 6$ into $\{1,2\}, \{3,4\}$ and $\{5,6\}$. The pictographs show the large size pattern of the corresponding states.
        }
        \label{fig:MAXMINCut}
\end{figure}

\paragraph{Antiferromagnetic limit:}
For $\mu/\sigma \to \infty$ the minimal constraint problem can be connected to the \textsc{MaxCut} problem on the complete graph $K_n$. In this case all interactions are antiferromagnetic, and can be set to $J_{ij} = 1$  due to the rescaling property mentioned above.
Assuming $n = |V|$ to be an even number of vertices, then the maximal cut is given by $(n/2)^2$ and thus, the lowest eigenvalue of the corresponding Ising Hamiltonian is given by $l_0 = -n/2$ and has a gap of $2$ i.e.\@ independent of $n$. 
For the instructive example of $n=3k$  ($k\in\mathbb{N}$), the minimal single violator energy is given by $a_1 = -\frac{n}{2}\left(1+\frac{n}{3}\right)$.
Then the lower bounds $c_{-i}$ can be analytically found
where the largest is given by
\begin{equation}    c_{-1} = \frac{n^2}{6} + \left\{\begin{array}{rl}
         2&,  \text{ if } n = 3 k \text{ for some } k\in \mathbb{N}\\
         \frac{4}{3}&, \text{ else} 
    \end{array}\right. .
\end{equation}
The scaling of $c=c_{-1}$ is therefore $\Theta(n^2)$ i.e.\@ \emph{quadratic} in $n$.
For a graphical representation of the relevant states see Fig.~\ref{fig:MAXMINCut}.\\

\paragraph{Ferromagnetic limit:} 
The limit  $\mu/\sigma \to -\infty$ is reached when setting $J_{ij} = -1$ i.e.\@ all pair interactions are ferromagnetic. In this case, the lowest eigenvalue is given by $l_0 = -n(n-1)/2$ and shows a gap of $2(n-1)$. In the parity picture, violating a single constraint can be done at minimal cost of a single spinflip. Starting from the ground state in $\mathcal{S}_0$ flipping the spin $(1,n)$ results in a state from $\mathcal{S}_{[1,n-1]}$ with energy $l_0+2$ and thus $c = c_{-1} = 2n - 4$. The scaling of $c$ is therefore $\Theta(n)$ i.e.\@ \emph{linear} in $n$.\\

\paragraph{\textsc{SK-SpinGlass}:} 
\label{sec:AnalyticalResults:SK}

Of special interest is the case $\mu/ \sigma = 0$.
The numerical results for $\overline{c_{-1}}$ as function of $n$ are shown in Figure~\ref{fig:Slopes_mu_over_sigma}. The following arguments suggest that in the Gaussian case the scaling of $\overline{c_{-1}}$ goes as $\sqrt{n}\log(n)$. 

To begin with, we note, that for every fixed spin configuration $\vec{\sigma}_z\in\{-1,1\}^n$ the eigenvalues of $H_\mathrm{logic}(\vec{\sigma}_z)$ can be seen as a random variable w.r.t.\@\@ the distribution determining the interaction strengths. If this distribution is Gaussian, then - according to the central limit theorem - the eigenvalues are also normally distributed with variance $\sigma^2 = n(n-1)/2$.

However, the $2^n$ different eigenvalues are clearly not independent.
They are strongly correlated with covariance matrix \cite{Panchenko_2012}
\begin{equation}
\label{eq:Correlations}
    \mathbb{E}(H_\mathrm{logic}(\vec{\sigma}_z)H_\mathrm{logic}(\vec{\tau}_z)) = \frac{1}{2}\left(\sum_{i=1}^n \sigma_z^i\tau_z^i\right)^2 -\frac{n}{2}.
\end{equation}
Note, that finding the minimum $\min_{\vec{\sigma}_z} H_\mathrm{logic}(\vec{\sigma}_z)$ is challenging due to the presents of these correlations. 

In the case of independent random variables, the limiting order statistics can be classified into one of three universally classes according to extreme value theory \cite{KotzNadarajah}. Moreover, for $m$ Gaussian variables $\{G_1,...,G_m\}$ the limiting distribution of their minimum $M = \min(G_1,..,G_m)$ is given by the Gumbel distribution
$\mathrm{Gumbel}(\alpha,\beta)$ with parameters 
\begin{equation}
\label{eq:GumbelParameters}
    \alpha = F^{-1}\left(1-\frac{1}{m}\right),\quad
    \beta = F^{-1}\left(1-\frac{1}{e m}\right) -\alpha.
\end{equation}
Here $e$ is the Euler constant and $F^{-1}$ denotes the quantil function, which in the case of standard normal variables is given by the probit $\sqrt{2}\mathrm{erf}^{-1}(2p-1)$. By fixing the distribution, the parameters $\alpha$ and $\beta$ only depend on the number of variables $m$. We want to emphasise, that setting $m = 2^n$ models all eigenvalues as independent. 
Since the factor $\sigma^{-1}$ normalizes each eigenvalue, $\overline{M}_{\mathrm{ind}} := -\sigma (\alpha + \Gamma \beta)$ is the expected minimal energy in the independent case ($\Gamma$ is the Euler-Mascheroni constant). 

Numerically we see, that choosing the eigenvalues to be independent, results (on average) in too small energies $\overline{l_0} > \overline{M}_{\mathrm{ind}} (m= 2^{n})$.
However, we observe that by introducing a free parameter $\delta < 1$ in order to 
decrease the number of realizations $m = 2^{n\delta}$ gives a reasonably good approximation 
$\overline{l_0} \approx \overline{M}_{\mathrm{ind}} (m= 2^{n\delta})$ for $\delta \approx 0.798158(4)$ [cf.\@Fig.~\ref{fig:SK_ind_EVT}] \footnote{
By curiosity, we note that value is really close $\sqrt{2/\pi} = 0.797884...\,$. The value $\sqrt{2/\pi}$ can be related to a normal distribution: If $X$ is a standard normal variable, then
$|X|$ defines another random variable with mean $\sqrt{2/\pi}$.
}.
If one allows $\delta$ to be a function of $n$ then the statement is trivial, but interestingly a constant $\delta$ gives rise to a good approximation.
To which extend our analysis captures the large $n$ behaviour is outside the scope of the present work. However, as we show in the appendix $\overline{M}_{\mathrm{ind}} \approx -\sqrt{\delta \log(2) } n^{\frac{3}{2}}$ for large $n$. Since 
$\sqrt{\delta \log(2)} \approx 0.743$ this is in accordance with Parisis result \cite{Parisi1980ASO,Talagrand}
\begin{equation}
    \label{eq:Parisi}
    \overline{l_0} = (-0.763167... + o(1))n^{\frac{3}{2}}.
\end{equation}
On the other hand, finding $a_1$ corresponds to finding the groundstate of the physical Hamiltonian $H_{\mathrm{phys}}$ restricted to the subspace of single violator states. Finding the single violator state w.r.t.\@ the plaquette indexed by $[k,l]$ is equivalent to finding the ground state
of $H_{[k,l]} := \sum J'_{ij}\sigma_z^{i}\sigma_z^j$ [cf.\@ Fig.~\ref{fig:Parity}(c)], with
\begin{equation}
\label{eq:S1_to_n2Ham_Reduction}
  J'_{ij} := \left\{
  \begin{array}{rl}
      -J_{ij}  & \text{ if } i \leq k \text{ and } j > l  \\
      J_{ij} & \text{ else}
  \end{array}
  \right..  
\end{equation}
Hence, minimizing over all plaquettes gives the minimal single-violator energy $a_1$. 
Due to symmetry, $-J_{ij}$ is again a standard normal Gaussian variable and therefore the eigenenergies associated to single violators are distributed according to $N(0,\sigma^2)$. Similar to Eq.~\eqref{eq:Correlations} these eigenstates are highly correlated, but nevertheless we find a numerically well justified approximation
$\overline{a_1} \approx \overline{M}_{\mathrm{ind}} (m= 2^{n\delta} n(n+1)/12)$, with $\delta$ as above. 
Here, the quadratic terms count the average number of spins that have to be flipped to induce a parity defect according to Fig.~\ref{fig:Parity}(c).
As we show in appendix~\ref{sec:Appendix_SKModel}, the energy difference between single violator ground states and logical ground state $\overline{l_0-a_1}$ scales to leading order in $n$ as
\begin{equation}
    f_1(n):=
    \frac{1}{2\sqrt{\delta \log(2)}} \sqrt{n}\log\left[ \frac{n(n+1)}{12} \right].
    \label{eq:f1}
\end{equation}
Since the gap $\Delta_l$ can be neglected for large $n$, the scaling of $\overline{c_{-1}}\propto \overline{l_0-a_1}$
[cf.\@ Fig.~\ref{fig:SK_ind_EVT}]. 

Similar arguments can be used for the scaling analysis of the remaining lower bounds $\frac{1}{k}(l_0 - a_k)$. Here, the subspace $\mathcal{S}_k$ spanned by states with $k$ parity constraints violated has $\mathcal{O}(n^{2k})$ elements. Setting $p_k(n)$ to a polynomial of order $2k$ and modelling $\overline{a_k}$ as $2^{\delta n}p_k(n)$ i.i.d.\@ Gaussian variables results in a scaling of 
$c_{-k}$ as $\frac{1}{k}\sqrt{n}\log(n^{2k}) \propto \sqrt{n}\log(n)$ analog to Eq.~\eqref{eq:f1}. \\

\paragraph{General case:}

Finally, let us discuss the case when $J_{ij}$ are neither centered distributed random variables nor close to the limits $\mu/\sigma = \pm \infty$. 
In general the bound $c_{-1}$ depends on the lowest eigenvalue $l_0$, the gap $\Delta_1$ and the smallest single violator energy $a_1$.

If $\mu \neq 0$, the distributions of eigenenergies of the logical Hamiltonian are shifted in contrast to the previously considered case of $\mu = 0$. As an example we consider in the following Gaussian distributed couplings $J_{ij}\sim N(\mu,1)$. Our argument builds on the fact that two normal variables with parameters $(\mu_1,\sigma_1^2)$ and $(\mu_2,\sigma_2^2)$
add up to a single normal variable
$N(\mu_1+\mu_2, \sigma_1^2+\sigma_2^2)$.
Thus, the eigenvalue corresponding to the all-ones logical state $(1,...,1)$ is distributed via $N(\mu m,m)$, where $m$ denotes the number of physical spins. Likewise, the distribution of a state with equally many $-1$ and $+1$ is centered around zero $\sim N(0,m)$.
More generally, for fixed $k$, there are $2\binom{n}{k}$ combinations of eigenvalues, where every combination is distributed according to a Gaussian with mean 
\begin{equation}
    \mu_k = \mu \frac{(n-2k)^2-n}{2}
\end{equation}
and variance $m$.
This 'splitting' is the reason for the different behaviour of the smallest eigenvalue in the two cases $\mu > 0$ and $\mu<0$. 

In the case $\mu > 0$, the probability that the lowest energies originate from one of the $2\binom{2}{n/2}$ states centered around zero is largest.
Since these are exponentially many states, we assume the expectation value for the ground state energy to be similar to the $\mu = 0$ case.
The single violator ground state for the limit $\mu/\sigma \to \infty$, denoted by  $\phi_{\infty}$  [cf. Fig.~\ref{fig:MAXMINCut}(bottom)],
is Gaussian distributed w.r.t.\@ $J_{ij}$ with mean $-\mu n^2(\frac{1}{6} + \mathcal{O}(n^{-1}))$ and variance $m$. 
Therefore, the single violator ground-state energy lies on average quadratically deeper than the logical ground state energy. 

On the contrary, for $\mu < 0$, the larger $|\mu|$ is, the more likely it is, that the smallest eigenvalue is one of the ferromagnetic states $(1,...,1)$ or $(-1,...,-1)$.
As shown, in the total ferromagnetic case, the gap $\Delta_l$ grows linearly with $n$. Since $l_0-a_1 = 2$ this linear increase is the main contribution to $c_{-1}$. 
In the large-$n$ limit we expect that the gap scales linearly. Since our numerical simulation is limited up to $n=25$, we see this linear behaviour only if $|\mu|$ is large enough. As one can further see in Fig.~\ref{fig:Slopes_mu_over_sigma} as $\mu$ gets smaller, the scaling exponent drops due to the fact that the expected difference $l_0-a_1$ tends to get smaller for more negative $\mu$. However, by further increasing $|\mu|$ the linear scaling of the gap becomes apparent.

\section{Discussion}

We have shown numerically and analytically the scaling of the minimal constraint strengths in the parity based encoding for various classes of optimization problems. In the parity scheme, the optimization problem is encoded in the local fields only, and thus the classes of problems differ only in the statistics of the local fields. We have shown that the large size scaling is mainly determined by the sign of the bias $\mu$. The observed differences stem from two distinct effects:
(a) linear growth of the gap and (b) quadratic deep lying single violator states.
While the scaling of the gap is the important factor for problems with
predominantly negative couplings, the properties of the single violator states govern the regime with predominately positive couplings. Thus, the large size scaling of the expected constraint strengths grows linearly or quadratically with the system size in the respective cases.

One special point in between these regimes arises when the couplings are centered with mean zero. Finally, a non-rigorous analysis suggests a sub-linear behaviour given by $\sqrt{n}\log(n)$. 
We want to emphasize, that the minimization of the constraint strengths could have direct influence on the minimal gap during AQO. In the QAOA setting, the constraint values could serve as additional variational parameters. In the latter case the optimized values can serve as good initial choice of these parameters.

\section*{Acknowledgement}
The authors thank Kilian Ender and Clemens Dlaska for the useful discussions and comments on the paper. Work was supported by the Austrian Science Fund (FWF) through a START grant under Project No. Y1067-N27 and the SFB BeyondC Project No. F7108-N38, the Hauser-Raspe foundation, and the European Union's Horizon 2020 research and innovation program under grant agreement No. 817482. This material is based upon work supported by the Defense Advanced Research Projects Agency (DARPA) under Contract No. HR001120C0068. Any opinions, findings and conclusions or recommendations expressed in this material are those of the author(s) and do not necessarily reflect the views of DARPA.


\begin{thebibliography}{33}%
\makeatletter
\providecommand \@ifxundefined [1]{%
 \@ifx{#1\undefined}
}%
\providecommand \@ifnum [1]{%
 \ifnum #1\expandafter \@firstoftwo
 \else \expandafter \@secondoftwo
 \fi
}%
\providecommand \@ifx [1]{%
 \ifx #1\expandafter \@firstoftwo
 \else \expandafter \@secondoftwo
 \fi
}%
\providecommand \natexlab [1]{#1}%
\providecommand \enquote  [1]{``#1''}%
\providecommand \bibnamefont  [1]{#1}%
\providecommand \bibfnamefont [1]{#1}%
\providecommand \citenamefont [1]{#1}%
\providecommand \href@noop [0]{\@secondoftwo}%
\providecommand \href [0]{\begingroup \@sanitize@url \@href}%
\providecommand \@href[1]{\@@startlink{#1}\@@href}%
\providecommand \@@href[1]{\endgroup#1\@@endlink}%
\providecommand \@sanitize@url [0]{\catcode `\\12\catcode `\$12\catcode
  `\&12\catcode `\#12\catcode `\^12\catcode `\_12\catcode `\%12\relax}%
\providecommand \@@startlink[1]{}%
\providecommand \@@endlink[0]{}%
\providecommand \url  [0]{\begingroup\@sanitize@url \@url }%
\providecommand \@url [1]{\endgroup\@href {#1}{\urlprefix }}%
\providecommand \urlprefix  [0]{URL }%
\providecommand \Eprint [0]{\href }%
\providecommand \doibase [0]{http://dx.doi.org/}%
\providecommand \selectlanguage [0]{\@gobble}%
\providecommand \bibinfo  [0]{\@secondoftwo}%
\providecommand \bibfield  [0]{\@secondoftwo}%
\providecommand \translation [1]{[#1]}%
\providecommand \BibitemOpen [0]{}%
\providecommand \bibitemStop [0]{}%
\providecommand \bibitemNoStop [0]{.\EOS\space}%
\providecommand \EOS [0]{\spacefactor3000\relax}%
\providecommand \BibitemShut  [1]{\csname bibitem#1\endcsname}%
\let\auto@bib@innerbib\@empty
\bibitem [{\citenamefont {Lucas}(2014)}]{Lucas}%
  \BibitemOpen
  \bibfield  {author} {\bibinfo {author} {\bibfnamefont {A.}~\bibnamefont
  {Lucas}},\ }\href {\doibase 10.3389/fphy.2014.00005} {\bibfield  {journal}
  {\bibinfo  {journal} {Frontiers in Physics}\ }\textbf {\bibinfo {volume}
  {2}},\ \bibinfo {pages} {5} (\bibinfo {year} {2014})}\BibitemShut {NoStop}%
\bibitem [{\citenamefont {Farhi}\ \emph {et~al.}(2014)\citenamefont {Farhi},
  \citenamefont {Goldstone},\ and\ \citenamefont {Gutmann}}]{FarhiQAOA}%
  \BibitemOpen
  \bibfield  {author} {\bibinfo {author} {\bibfnamefont {E.}~\bibnamefont
  {Farhi}}, \bibinfo {author} {\bibfnamefont {J.}~\bibnamefont {Goldstone}}, \
  and\ \bibinfo {author} {\bibfnamefont {S.}~\bibnamefont {Gutmann}},\
  }\href@noop {} {\enquote {\bibinfo {title} {A quantum approximate
  optimization algorithm},}\ } (\bibinfo {year} {2014}),\ \Eprint
  {http://arxiv.org/abs/1411.4028} {arXiv:1411.4028 [quant-ph]} \BibitemShut
  {NoStop}%
\bibitem [{\citenamefont {Farhi}\ \emph {et~al.}(2000)\citenamefont {Farhi},
  \citenamefont {Goldstone}, \citenamefont {Gutmann},\ and\ \citenamefont
  {Sipser}}]{FGGS}%
  \BibitemOpen
  \bibfield  {author} {\bibinfo {author} {\bibfnamefont {E.}~\bibnamefont
  {Farhi}}, \bibinfo {author} {\bibfnamefont {J.}~\bibnamefont {Goldstone}},
  \bibinfo {author} {\bibfnamefont {S.}~\bibnamefont {Gutmann}}, \ and\
  \bibinfo {author} {\bibfnamefont {M.}~\bibnamefont {Sipser}},\ }\href@noop {}
  {\  (\bibinfo {year} {2000})}\BibitemShut {NoStop}%
\bibitem [{\citenamefont {Susa}\ \emph {et~al.}(2018)\citenamefont {Susa},
  \citenamefont {Yamashiro}, \citenamefont {Yamamoto}, \citenamefont {Hen},
  \citenamefont {Lidar},\ and\ \citenamefont {Nishimori}}]{SusaNishimori}%
  \BibitemOpen
  \bibfield  {author} {\bibinfo {author} {\bibfnamefont {Y.}~\bibnamefont
  {Susa}}, \bibinfo {author} {\bibfnamefont {Y.}~\bibnamefont {Yamashiro}},
  \bibinfo {author} {\bibfnamefont {M.}~\bibnamefont {Yamamoto}}, \bibinfo
  {author} {\bibfnamefont {I.}~\bibnamefont {Hen}}, \bibinfo {author}
  {\bibfnamefont {D.~A.}\ \bibnamefont {Lidar}}, \ and\ \bibinfo {author}
  {\bibfnamefont {H.}~\bibnamefont {Nishimori}},\ }\href {\doibase
  10.1103/PhysRevA.98.042326} {\bibfield  {journal} {\bibinfo  {journal} {Phys.
  Rev. A}\ }\textbf {\bibinfo {volume} {98}},\ \bibinfo {pages} {042326}
  (\bibinfo {year} {2018})}\BibitemShut {NoStop}%
\bibitem [{\citenamefont {Kadowaki}\ and\ \citenamefont
  {Nishimori}(1998)}]{Kadowaki1998QA}%
  \BibitemOpen
  \bibfield  {author} {\bibinfo {author} {\bibfnamefont {T.}~\bibnamefont
  {Kadowaki}}\ and\ \bibinfo {author} {\bibfnamefont {H.}~\bibnamefont
  {Nishimori}},\ }\href {\doibase 10.1103/physreve.58.5355} {\bibfield
  {journal} {\bibinfo  {journal} {Physical Review E}\ }\textbf {\bibinfo
  {volume} {58}},\ \bibinfo {pages} {5355–5363} (\bibinfo {year}
  {1998})}\BibitemShut {NoStop}%
\bibitem [{\citenamefont {Kadowaki}(2002)}]{Kadowaki2002Thesis}%
  \BibitemOpen
  \bibfield  {author} {\bibinfo {author} {\bibfnamefont {T.}~\bibnamefont
  {Kadowaki}},\ }\href@noop {} {\enquote {\bibinfo {title} {Study of
  optimization problems by quantum annealing},}\ } (\bibinfo {year} {2002}),\
  \Eprint {http://arxiv.org/abs/quant-ph/0205020} {arXiv:quant-ph/0205020
  [quant-ph]} \BibitemShut {NoStop}%
\bibitem [{\citenamefont {Albash}\ and\ \citenamefont {Lidar}(2018)}]{AlbLid}%
  \BibitemOpen
  \bibfield  {author} {\bibinfo {author} {\bibfnamefont {T.}~\bibnamefont
  {Albash}}\ and\ \bibinfo {author} {\bibfnamefont {D.~A.}\ \bibnamefont
  {Lidar}},\ }\href {\doibase 10.1103/RevModPhys.90.015002} {\bibfield
  {journal} {\bibinfo  {journal} {Rev. Mod. Phys.}\ }\textbf {\bibinfo {volume}
  {90}},\ \bibinfo {pages} {015002} (\bibinfo {year} {2018})}\BibitemShut
  {NoStop}%
\bibitem [{\citenamefont {Hauke}\ \emph {et~al.}(2020)\citenamefont {Hauke},
  \citenamefont {Katzgraber}, \citenamefont {Lechner}, \citenamefont
  {Nishimori},\ and\ \citenamefont {Oliver}}]{Hauke2020PerspectivesOQ}%
  \BibitemOpen
  \bibfield  {author} {\bibinfo {author} {\bibfnamefont {P.}~\bibnamefont
  {Hauke}}, \bibinfo {author} {\bibfnamefont {H.~G.}\ \bibnamefont
  {Katzgraber}}, \bibinfo {author} {\bibfnamefont {W.}~\bibnamefont {Lechner}},
  \bibinfo {author} {\bibfnamefont {H.}~\bibnamefont {Nishimori}}, \ and\
  \bibinfo {author} {\bibfnamefont {W.~D.}\ \bibnamefont {Oliver}},\
  }\href@noop {} {\bibfield  {journal} {\bibinfo  {journal} {Reports on
  progress in physics. Physical Society}\ } (\bibinfo {year}
  {2020})}\BibitemShut {NoStop}%
\bibitem [{\citenamefont {Katzgraber}\ \emph {et~al.}(2015)\citenamefont
  {Katzgraber}, \citenamefont {Hamze}, \citenamefont {Zhu}, \citenamefont
  {Ochoa},\ and\ \citenamefont {Munoz-Bauza}}]{Katzgraber2015SeekingQS}%
  \BibitemOpen
  \bibfield  {author} {\bibinfo {author} {\bibfnamefont {H.~G.}\ \bibnamefont
  {Katzgraber}}, \bibinfo {author} {\bibfnamefont {F.}~\bibnamefont {Hamze}},
  \bibinfo {author} {\bibfnamefont {Z.}~\bibnamefont {Zhu}}, \bibinfo {author}
  {\bibfnamefont {A.~J.}\ \bibnamefont {Ochoa}}, \ and\ \bibinfo {author}
  {\bibfnamefont {H.}~\bibnamefont {Munoz-Bauza}},\ }\href@noop {} {\bibfield
  {journal} {\bibinfo  {journal} {Physical Review X}\ }\textbf {\bibinfo
  {volume} {5}},\ \bibinfo {pages} {031026} (\bibinfo {year}
  {2015})}\BibitemShut {NoStop}%
\bibitem [{\citenamefont {Lidar}\ \emph {et~al.}(2014)\citenamefont {Lidar},
  \citenamefont {Wecker}, \citenamefont {Martinis}, \citenamefont {Job},
  \citenamefont {Troyer}, \citenamefont {Isakov}, \citenamefont {Boixo},
  \citenamefont {Wang} \emph {et~al.}}]{LidarTroyerGoogleQSU}%
  \BibitemOpen
  \bibfield  {author} {\bibinfo {author} {\bibfnamefont {D.~A.}\ \bibnamefont
  {Lidar}}, \bibinfo {author} {\bibfnamefont {D.}~\bibnamefont {Wecker}},
  \bibinfo {author} {\bibfnamefont {J.~M.}\ \bibnamefont {Martinis}}, \bibinfo
  {author} {\bibfnamefont {J.}~\bibnamefont {Job}}, \bibinfo {author}
  {\bibfnamefont {M.}~\bibnamefont {Troyer}}, \bibinfo {author} {\bibfnamefont
  {S.~V.}\ \bibnamefont {Isakov}}, \bibinfo {author} {\bibfnamefont
  {S.}~\bibnamefont {Boixo}}, \bibinfo {author} {\bibfnamefont
  {Z.}~\bibnamefont {Wang}},  \emph {et~al.},\ }\href@noop {} {\bibfield
  {journal} {\bibinfo  {journal} {Science}\ } (\bibinfo {year}
  {2014})}\BibitemShut {NoStop}%
\bibitem [{\citenamefont {Mbeng}\ \emph {et~al.}(2019)\citenamefont {Mbeng},
  \citenamefont {Privitera}, \citenamefont {Arceci},\ and\ \citenamefont
  {Santoro}}]{MbengSQA19}%
  \BibitemOpen
  \bibfield  {author} {\bibinfo {author} {\bibfnamefont {G.~B.}\ \bibnamefont
  {Mbeng}}, \bibinfo {author} {\bibfnamefont {L.}~\bibnamefont {Privitera}},
  \bibinfo {author} {\bibfnamefont {L.}~\bibnamefont {Arceci}}, \ and\ \bibinfo
  {author} {\bibfnamefont {G.~E.}\ \bibnamefont {Santoro}},\ }\href {\doibase
  10.1103/physrevb.99.064201} {\bibfield  {journal} {\bibinfo  {journal}
  {Physical Review B}\ }\textbf {\bibinfo {volume} {99}} (\bibinfo {year}
  {2019}),\ 10.1103/physrevb.99.064201}\BibitemShut {NoStop}%
\bibitem [{\citenamefont {Santoro}(2002)}]{SantoroAnnealing}%
  \BibitemOpen
  \bibfield  {author} {\bibinfo {author} {\bibfnamefont {G.~E.}\ \bibnamefont
  {Santoro}},\ }\href {\doibase 10.1126/science.1068774} {\bibfield  {journal}
  {\bibinfo  {journal} {Science}\ }\textbf {\bibinfo {volume} {295}},\ \bibinfo
  {pages} {2427–2430} (\bibinfo {year} {2002})}\BibitemShut {NoStop}%
\bibitem [{\citenamefont {Dickson}\ and\ \citenamefont {Amin}(2011)}]{Dick}%
  \BibitemOpen
  \bibfield  {author} {\bibinfo {author} {\bibfnamefont {N.~G.}\ \bibnamefont
  {Dickson}}\ and\ \bibinfo {author} {\bibfnamefont {M.~H.~S.}\ \bibnamefont
  {Amin}},\ }\href {\doibase 10.1103/PhysRevLett.106.050502} {\bibfield
  {journal} {\bibinfo  {journal} {Phys. Rev. Lett.}\ }\textbf {\bibinfo
  {volume} {106}},\ \bibinfo {pages} {050502} (\bibinfo {year}
  {2011})}\BibitemShut {NoStop}%
\bibitem [{\citenamefont {Hartmann}\ and\ \citenamefont
  {Lechner}(2019{\natexlab{a}})}]{HartmannCD}%
  \BibitemOpen
  \bibfield  {author} {\bibinfo {author} {\bibfnamefont {A.}~\bibnamefont
  {Hartmann}}\ and\ \bibinfo {author} {\bibfnamefont {W.}~\bibnamefont
  {Lechner}},\ }\href {\doibase 10.1088/1367-2630/ab14a0} {\bibfield  {journal}
  {\bibinfo  {journal} {New Journal of Physics}\ }\textbf {\bibinfo {volume}
  {21}},\ \bibinfo {pages} {043025} (\bibinfo {year}
  {2019}{\natexlab{a}})}\BibitemShut {NoStop}%
\bibitem [{\citenamefont {Hartmann}\ and\ \citenamefont
  {Lechner}(2019{\natexlab{b}})}]{HartmannIHD}%
  \BibitemOpen
  \bibfield  {author} {\bibinfo {author} {\bibfnamefont {A.}~\bibnamefont
  {Hartmann}}\ and\ \bibinfo {author} {\bibfnamefont {W.}~\bibnamefont
  {Lechner}},\ }\href {\doibase 10.1103/physreva.100.032110} {\bibfield
  {journal} {\bibinfo  {journal} {Physical Review A}\ }\textbf {\bibinfo
  {volume} {100}} (\bibinfo {year} {2019}{\natexlab{b}}),\
  10.1103/physreva.100.032110}\BibitemShut {NoStop}%
\bibitem [{\citenamefont {Zhou}\ \emph {et~al.}(2020)\citenamefont {Zhou},
  \citenamefont {Wang}, \citenamefont {Choi}, \citenamefont {Pichler},\ and\
  \citenamefont {Lukin}}]{ZhouPichlerLukin2020QAOA}%
  \BibitemOpen
  \bibfield  {author} {\bibinfo {author} {\bibfnamefont {L.}~\bibnamefont
  {Zhou}}, \bibinfo {author} {\bibfnamefont {S.-T.}\ \bibnamefont {Wang}},
  \bibinfo {author} {\bibfnamefont {S.}~\bibnamefont {Choi}}, \bibinfo {author}
  {\bibfnamefont {H.}~\bibnamefont {Pichler}}, \ and\ \bibinfo {author}
  {\bibfnamefont {M.~D.}\ \bibnamefont {Lukin}},\ }\href {\doibase
  10.1103/physrevx.10.021067} {\bibfield  {journal} {\bibinfo  {journal}
  {Physical Review X}\ }\textbf {\bibinfo {volume} {10}} (\bibinfo {year}
  {2020}),\ 10.1103/physrevx.10.021067}\BibitemShut {NoStop}%
\bibitem [{\citenamefont {Farhi}\ and\ \citenamefont
  {Harrow}(2016)}]{Farhi2016QSuprem}%
  \BibitemOpen
  \bibfield  {author} {\bibinfo {author} {\bibfnamefont {E.}~\bibnamefont
  {Farhi}}\ and\ \bibinfo {author} {\bibfnamefont {A.~W.}\ \bibnamefont
  {Harrow}},\ }\href@noop {} {\enquote {\bibinfo {title} {Quantum supremacy
  through the quantum approximate optimization algorithm},}\ } (\bibinfo {year}
  {2016}),\ \Eprint {http://arxiv.org/abs/1602.07674} {arXiv:1602.07674
  [quant-ph]} \BibitemShut {NoStop}%
\bibitem [{\citenamefont {Wecker}\ \emph {et~al.}(2016)\citenamefont {Wecker},
  \citenamefont {Hastings},\ and\ \citenamefont {Troyer}}]{Wecker2016}%
  \BibitemOpen
  \bibfield  {author} {\bibinfo {author} {\bibfnamefont {D.}~\bibnamefont
  {Wecker}}, \bibinfo {author} {\bibfnamefont {M.~B.}\ \bibnamefont
  {Hastings}}, \ and\ \bibinfo {author} {\bibfnamefont {M.}~\bibnamefont
  {Troyer}},\ }\href {\doibase 10.1103/physreva.94.022309} {\bibfield
  {journal} {\bibinfo  {journal} {Physical Review A}\ }\textbf {\bibinfo
  {volume} {94}} (\bibinfo {year} {2016}),\
  10.1103/physreva.94.022309}\BibitemShut {NoStop}%
\bibitem [{\citenamefont {Choi}(2008)}]{Choi2008ME}%
  \BibitemOpen
  \bibfield  {author} {\bibinfo {author} {\bibfnamefont {V.}~\bibnamefont
  {Choi}},\ }\href {\doibase 10.1007/s11128-008-0082-9} {\bibfield  {journal}
  {\bibinfo  {journal} {Quantum Information Processing}\ }\textbf {\bibinfo
  {volume} {7}},\ \bibinfo {pages} {193–209} (\bibinfo {year}
  {2008})}\BibitemShut {NoStop}%
\bibitem [{\citenamefont {Choi}(2010)}]{Choi2010ME}%
  \BibitemOpen
  \bibfield  {author} {\bibinfo {author} {\bibfnamefont {V.}~\bibnamefont
  {Choi}},\ }\href {\doibase 10.1007/s11128-010-0200-3} {\bibfield  {journal}
  {\bibinfo  {journal} {Quantum Information Processing}\ }\textbf {\bibinfo
  {volume} {10}},\ \bibinfo {pages} {343–353} (\bibinfo {year}
  {2010})}\BibitemShut {NoStop}%
\bibitem [{\citenamefont {Bunyk}\ \emph {et~al.}(2014)\citenamefont {Bunyk},
  \citenamefont {Hoskinson}, \citenamefont {Johnson}, \citenamefont
  {Tolkacheva}, \citenamefont {Altomare}, \citenamefont {Berkley},
  \citenamefont {Harris}, \citenamefont {Hilton}, \citenamefont {Lanting},
  \citenamefont {Przybysz},\ and\ \citenamefont {et~al.}}]{Bunyk2014Chimera}%
  \BibitemOpen
  \bibfield  {author} {\bibinfo {author} {\bibfnamefont {P.~I.}\ \bibnamefont
  {Bunyk}}, \bibinfo {author} {\bibfnamefont {E.~M.}\ \bibnamefont
  {Hoskinson}}, \bibinfo {author} {\bibfnamefont {M.~W.}\ \bibnamefont
  {Johnson}}, \bibinfo {author} {\bibfnamefont {E.}~\bibnamefont {Tolkacheva}},
  \bibinfo {author} {\bibfnamefont {F.}~\bibnamefont {Altomare}}, \bibinfo
  {author} {\bibfnamefont {A.~J.}\ \bibnamefont {Berkley}}, \bibinfo {author}
  {\bibfnamefont {R.}~\bibnamefont {Harris}}, \bibinfo {author} {\bibfnamefont
  {J.~P.}\ \bibnamefont {Hilton}}, \bibinfo {author} {\bibfnamefont
  {T.}~\bibnamefont {Lanting}}, \bibinfo {author} {\bibfnamefont {A.~J.}\
  \bibnamefont {Przybysz}}, \ and\ \bibinfo {author} {\bibnamefont {et~al.}},\
  }\href {\doibase 10.1109/tasc.2014.2318294} {\bibfield  {journal} {\bibinfo
  {journal} {IEEE Transactions on Applied Superconductivity}\ }\textbf
  {\bibinfo {volume} {24}},\ \bibinfo {pages} {1–10} (\bibinfo {year}
  {2014})}\BibitemShut {NoStop}%
\bibitem [{\citenamefont {Lechner}\ \emph {et~al.}(2015)\citenamefont
  {Lechner}, \citenamefont {Hauke},\ and\ \citenamefont {Zoller}}]{LHZ}%
  \BibitemOpen
  \bibfield  {author} {\bibinfo {author} {\bibfnamefont {W.}~\bibnamefont
  {Lechner}}, \bibinfo {author} {\bibfnamefont {P.}~\bibnamefont {Hauke}}, \
  and\ \bibinfo {author} {\bibfnamefont {P.}~\bibnamefont {Zoller}},\ }\href
  {\doibase 10.1126/sciadv.1500838} {\bibfield  {journal} {\bibinfo  {journal}
  {Science Advances}\ }\textbf {\bibinfo {volume} {1}} (\bibinfo {year}
  {2015}),\ 10.1126/sciadv.1500838}\BibitemShut {NoStop}%
\bibitem [{\citenamefont {Lechner}(2018)}]{LechnerQAOA}%
  \BibitemOpen
  \bibfield  {author} {\bibinfo {author} {\bibfnamefont {W.}~\bibnamefont
  {Lechner}},\ }\href@noop {} {\enquote {\bibinfo {title} {Quantum approximate
  optimization with parallelizable gates},}\ } (\bibinfo {year} {2018}),\
  \Eprint {http://arxiv.org/abs/1802.01157} {arXiv:1802.01157 [quant-ph]}
  \BibitemShut {NoStop}%
\bibitem [{\citenamefont {Glaetzle}\ \emph {et~al.}(2017)\citenamefont
  {Glaetzle}, \citenamefont {van Bijnen}, \citenamefont {Zoller},\ and\
  \citenamefont {Lechner}}]{GvBZL}%
  \BibitemOpen
  \bibfield  {author} {\bibinfo {author} {\bibfnamefont {A.~W.}\ \bibnamefont
  {Glaetzle}}, \bibinfo {author} {\bibfnamefont {R.~M.~W.}\ \bibnamefont {van
  Bijnen}}, \bibinfo {author} {\bibfnamefont {P.}~\bibnamefont {Zoller}}, \
  and\ \bibinfo {author} {\bibfnamefont {W.}~\bibnamefont {Lechner}},\ }\href
  {https://doi.org/10.1038/ncomms15813} {\bibfield  {journal} {\bibinfo
  {journal} {Nature Communications}\ }\textbf {\bibinfo {volume} {8}},\
  \bibinfo {pages} {15813 EP } (\bibinfo {year} {2017})}\BibitemShut {NoStop}%
\bibitem [{\citenamefont {Leib}\ \emph {et~al.}(2016)\citenamefont {Leib},
  \citenamefont {Zoller},\ and\ \citenamefont {Lechner}}]{Leib2016Transmon}%
  \BibitemOpen
  \bibfield  {author} {\bibinfo {author} {\bibfnamefont {M.}~\bibnamefont
  {Leib}}, \bibinfo {author} {\bibfnamefont {P.}~\bibnamefont {Zoller}}, \ and\
  \bibinfo {author} {\bibfnamefont {W.}~\bibnamefont {Lechner}},\ }\href@noop
  {} {\bibfield  {journal} {\bibinfo  {journal} {Quantum Science and
  Technology}\ }\textbf {\bibinfo {volume} {1}},\ \bibinfo {pages} {015008}
  (\bibinfo {year} {2016})}\BibitemShut {NoStop}%
\bibitem [{\citenamefont {Albash}\ \emph {et~al.}(2016)\citenamefont {Albash},
  \citenamefont {Vinci},\ and\ \citenamefont {Lidar}}]{AlbLidSQA}%
  \BibitemOpen
  \bibfield  {author} {\bibinfo {author} {\bibfnamefont {T.}~\bibnamefont
  {Albash}}, \bibinfo {author} {\bibfnamefont {W.}~\bibnamefont {Vinci}}, \
  and\ \bibinfo {author} {\bibfnamefont {D.~A.}\ \bibnamefont {Lidar}},\ }\href
  {\doibase 10.1103/PhysRevA.94.022327} {\bibfield  {journal} {\bibinfo
  {journal} {Phys. Rev. A}\ }\textbf {\bibinfo {volume} {94}},\ \bibinfo
  {pages} {022327} (\bibinfo {year} {2016})}\BibitemShut {NoStop}%
\bibitem [{\citenamefont {Goemans}\ and\ \citenamefont
  {Williamson}(1995)}]{GoemansWilliamson1995improved}%
  \BibitemOpen
  \bibfield  {author} {\bibinfo {author} {\bibfnamefont {M.~X.}\ \bibnamefont
  {Goemans}}\ and\ \bibinfo {author} {\bibfnamefont {D.~P.}\ \bibnamefont
  {Williamson}},\ }\href@noop {} {\bibfield  {journal} {\bibinfo  {journal}
  {Journal of the ACM (JACM)}\ }\textbf {\bibinfo {volume} {42}},\ \bibinfo
  {pages} {1115} (\bibinfo {year} {1995})}\BibitemShut {NoStop}%
\bibitem [{\citenamefont {Panchenko}(2012)}]{Panchenko_2012}%
  \BibitemOpen
  \bibfield  {author} {\bibinfo {author} {\bibfnamefont {D.}~\bibnamefont
  {Panchenko}},\ }\href {\doibase 10.1007/s10955-012-0586-7} {\bibfield
  {journal} {\bibinfo  {journal} {Journal of Statistical Physics}\ }\textbf
  {\bibinfo {volume} {149}},\ \bibinfo {pages} {362–383} (\bibinfo {year}
  {2012})}\BibitemShut {NoStop}%
\bibitem [{\citenamefont {Kotz}\ and\ \citenamefont
  {Nadarajah}(2000)}]{KotzNadarajah}%
  \BibitemOpen
  \bibfield  {author} {\bibinfo {author} {\bibfnamefont {S.}~\bibnamefont
  {Kotz}}\ and\ \bibinfo {author} {\bibfnamefont {S.}~\bibnamefont
  {Nadarajah}},\ }\href {https://books.google.at/books?id=ZPW3CgAAQBAJ} {\emph
  {\bibinfo {title} {Extreme Value Distributions}}}\ (\bibinfo  {publisher}
  {World Scientific Publishing Company},\ \bibinfo {year} {2000})\BibitemShut
  {NoStop}%
\bibitem [{Note1()}]{Note1}%
  \BibitemOpen
  \bibinfo {note} {By curiosity, we note that value is really close $\protect
  \sqrt {2/\pi } = 0.797884...\protect \tmspace +\thinmuskip {.1667em}$. The
  value $\protect \sqrt {2/\pi }$ can be related to a normal distribution: If
  $X$ is a standard normal variable, then $|X|$ defines another random variable
  with mean $\protect \sqrt {2/\pi }$.}\BibitemShut {Stop}%
\bibitem [{\citenamefont {Parisi}(1980)}]{Parisi1980ASO}%
  \BibitemOpen
  \bibfield  {author} {\bibinfo {author} {\bibfnamefont {G.}~\bibnamefont
  {Parisi}}\ }(\bibinfo {year} {1980})\BibitemShut {NoStop}%
\bibitem [{\citenamefont {Talagrand}(2006)}]{Talagrand}%
  \BibitemOpen
  \bibfield  {author} {\bibinfo {author} {\bibfnamefont {M.}~\bibnamefont
  {Talagrand}},\ }\href@noop {} {\bibfield  {journal} {\bibinfo  {journal}
  {Annals of Mathematics}\ ,\ \bibinfo {pages} {221}} (\bibinfo {year}
  {2006})}\BibitemShut {NoStop}%
\bibitem [{\citenamefont {Strecok}(1968)}]{Strecok}%
  \BibitemOpen
  \bibfield  {author} {\bibinfo {author} {\bibfnamefont {A.~J.}\ \bibnamefont
  {Strecok}},\ }\href {http://www.jstor.org/stable/2004772} {\bibfield
  {journal} {\bibinfo  {journal} {Mathematics of Computation}\ }\textbf
  {\bibinfo {volume} {22}},\ \bibinfo {pages} {144} (\bibinfo {year}
  {1968})}\BibitemShut {NoStop}%
\end{thebibliography}

%

\appendix
\section{SK-model details}
\label{sec:Appendix_SKModel}

\begin{figure}
        \includegraphics[width =0.5\textwidth]{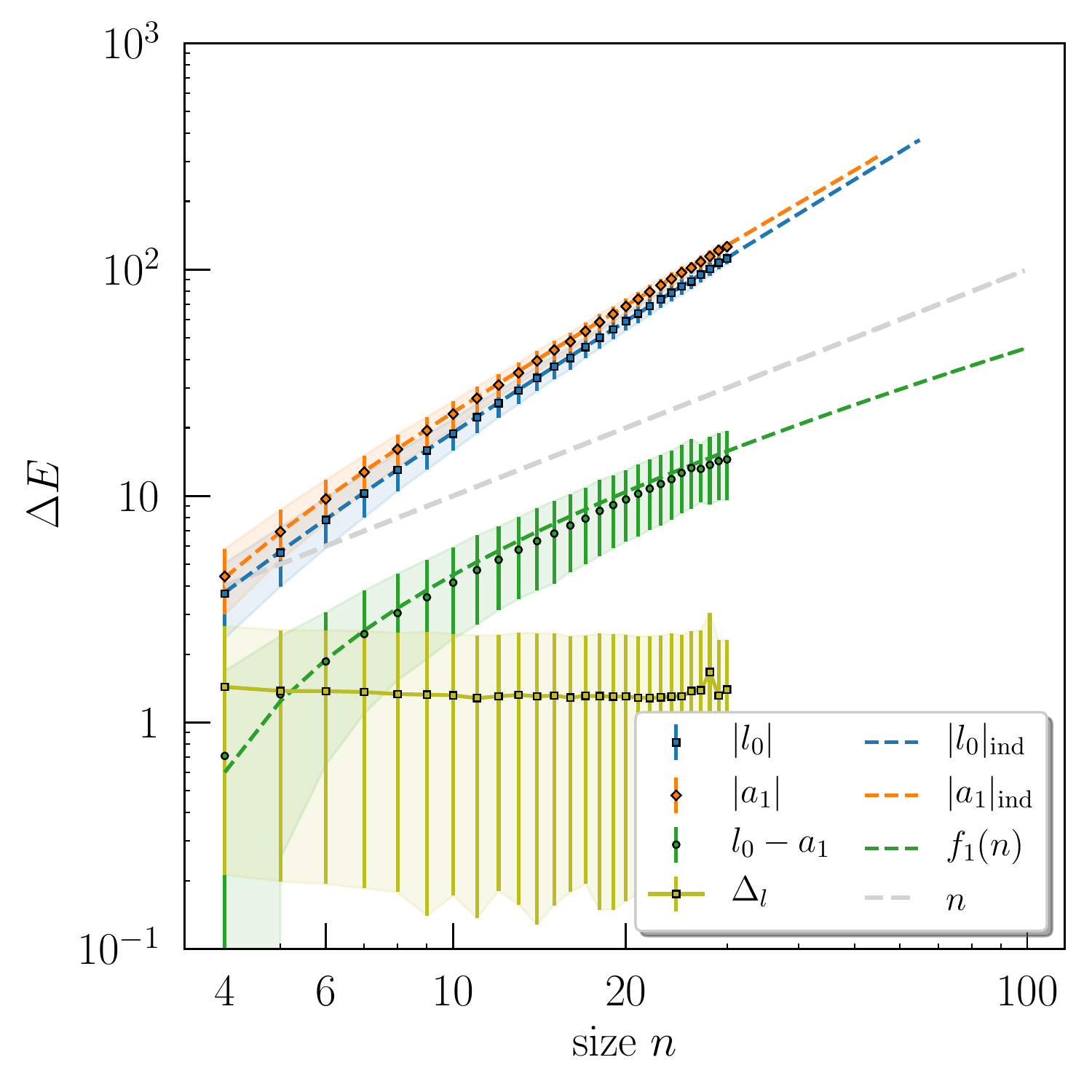}
        \caption[n]{Simulation of ground state energy $l_0$, minimal single violator energy $a_1$ and gap $\Delta_l$ for instances of the SK-Model. The corresponding number of instances drop from $10^5$ for $n=4$ to $32$ for $n=30$. 
        Beside the numerical results, we plot the approximations through our independent sampling idea $|l_0|_{\mathrm{ind}} = |\overline{M}_{\mathrm{ind}}(m = 2^{\delta n})|$, $|a_1|_{\mathrm{ind}}$ and $f_1$. 
        }
        \label{fig:SK_ind_EVT}
\end{figure}

The expected minimum of $m$  Gaussian variables is given as the mean of the corresponding Gumbel distribution via
$\overline{M_{\mathrm{ind}}} = - \sigma (\alpha + \Gamma \beta)$
with parameters as in Eq.~\eqref{eq:GumbelParameters}. Including the known 
quantil function for Gaussian variables one has
\begin{align}
    \label{eq:SK_Mean_ind}
    \overline{M_{\mathrm{ind}}}= \scalemath{0.9}{- \sqrt{n(n-1)}\bigg[ (1-\Gamma) \mathrm{erf}^{-1}\bigg(2\Big(1-\frac{1}{m}\Big)-1\bigg) }\\ \nonumber
    \scalemath{0.9}{
    + \Gamma \mathrm{erf}^{-1}\bigg(2\Big(1-\frac{1}{e m}\Big)-1\bigg) \bigg]
     } .
\end{align}
As described in the main text, we incorporate the correlations between the eigenvalues of the SK-Model by introducing a factor $\delta$ and observing 
$\overline{l_0} \approx \overline{M_{\mathrm{ind}}}(m = 2^{\delta n})$
for a fixed size independent $\delta$ [cf.\@ Fig.~\ref{fig:SK_ind_EVT}].

In order to motivate this simplified idea, we want to derive the large $n$
scaling and compare it to a well known result of Parisi~\cite{Parisi1980ASO}.
For $x$ close to one, the inverse error function can be approximated by $\mathrm{erf}^{-1}(x)\approx \sqrt{-\log (1-x^2)}$ \cite{Strecok}.
With $x = 2\big(1-2^{- \delta n}\big)-1$ 
we have $1-x^2 = (2^{\delta n} -1 )2^{2-2\delta n}$ and after decomposing the $\log$ of the product as the sum, we neglect the $-1$ over the $2^{\delta n}$ and arrive at
\begin{equation}
    \alpha \approx \sqrt{\left( \delta n - 2\right)\log(2)} .
\end{equation}
Doing the same for the second inverse error function term in Eq.~\eqref{eq:SK_Mean_ind} results into
\begin{equation}
    \beta + \alpha \approx \sqrt{\left( \delta n - 2\right)\log(2) + 1},
\end{equation}
where the additional $1$ comes from $\log(e) = 1$.
Neglecting this contribution gives $l_0 \approx - \sigma \alpha$
i.e.\@ 
\begin{equation}
    \label{eq:l0_approx_ind}
    \overline{l_0} \approx -\sqrt{n(n-1)}\sqrt{\left( \delta n - 2\right)\log(2)} .
\end{equation}
After expanding both square root expressions to leading order it holds
\begin{equation}
    \overline{l_0} \approx -\sqrt{\delta \log(2)} n^{\frac{2}{3}} \approx -0.7436\cdot  n^{\frac{3}{2}}.
\end{equation}
This motivates our approach, when compared to Parisis asymptotic  Eq.~\eqref{eq:Parisi}.

In the main text we have seen, finding $a_1$ corresponds to solve a family of ground state problems for Hamiltonians $H_{[k,l]}$, $k<l$ defined via via Eq.~\eqref{eq:S1_to_n2Ham_Reduction}. 
There are  $q = (n-2)(n-1)/2$ plaquettes and therefore $q$ possible ways to violate a single constraint. Since the standard Gaussian is centered and symmetric around zero,
each eigenvalue of these Hamiltonians are again distributed according $N(0,\sigma^2)$. Under the assumption: (i) the minimum of $2^{n\delta}$ independent variables corresponds to the expected lowest eigenvalue of the SK-model and  (ii) the eigenvalues of different $H_{[k,l]}$ are independent of each other one would expect $2^{\delta n} q$ independent random variables $\sim N(0,\sigma^2)$ can give us the expected smallest single violator energy. 
But assumption (ii) leads to way to low estimates for $\overline{a_1}$ since the $H_{[k,l]}$ are clearly not independent of each other. Following this road one gets a upper bound for  $\overline{l_0-a_1}$.
As described in the main text we found choosing 
$m = 2^{\delta n} p(n)$ Gaussian's $\sim N(0,\sigma^2)$ with the quadratic polynomial $p(n) = n(n+1)/12$ is well suited for modeling the expected single violator ground state energy [cf.\@ Fig.~\ref{fig:SK_ind_EVT} for a comparison with the numerical data].
Actually, the concrete form of $p(n)$ does not even play a role. Each (quadratic) polynomial leads to the same functional large $n$ behaviour for $\overline{l_0-a_1}$.

Setting $m = 2^{\delta n}p(n)$ in Eq.~\eqref{eq:SK_Mean_ind} and using the same approximations as before, one derives
\begin{equation}
    \label{eq:a1_approx_ind}
    \overline{a_1} \approx -\sqrt{n(n-1)}\sqrt{\left( \delta n - 2\right)\log(2) + \epsilon},
\end{equation}
with $\epsilon := \log(p(n))$.
Using $\sqrt{1+x} \approx 1+x/2$ for small $x$ to approximate the 
square root expressions in Eq.~\eqref{eq:l0_approx_ind} and Eq.~\eqref{eq:a1_approx_ind} leads to
\begin{equation}
    \overline{l_0}-\overline{a_1} \approx 
    \sqrt{n(n-1)}\frac{\epsilon}{2 \sqrt{\delta n \log(2)}} .
\end{equation}
Another Taylor expansion of $\sqrt{n(n-1)}$ reveals the leading order behaviour of $\Theta(\sqrt{n}\log(n))$ Eq.~ \eqref{eq:f1}.

\end{document}